\documentclass[aps,prl,twocolumn,showpacs,floatfix,amsmath,superscriptaddress]{revtex4}

\usepackage{bm}
\usepackage{graphicx}
\usepackage{latexsym}
\usepackage{amssymb}
\usepackage{color}

\begin{document}
\bibliographystyle{apsrev}

\title{Supersolid Vortex Crystals in Rydberg-dressed Bose-Einstein Condensates}

	\author{N. Henkel}
	\affiliation{Max Planck Institute for the Physics of Complex Systems, 01187 Dresden, Germany}

	\author{F. Cinti}
	\affiliation{Max Planck Institute for the Physics of Complex Systems, 01187 Dresden, Germany}
	\affiliation{Department of Physics, University of Alberta, Edmonton, Alberta, Canada}

	\author{P. Jain}
	\affiliation{Department of Physics, University of Alberta, Edmonton, Alberta, Canada}

	\author{G. Pupillo}	
	\affiliation{IQOQI and Institute for Theoretical Physics, University of Innsbruck, 6020 Innsbruck, Austria.}	
        \affiliation{ISIS (UMR 7006) and IPCMS (UMR 7504), Universit\'e de Strasbourg and CNRS, Strasbourg, France}

	\author{T. Pohl}		
	\affiliation{Max Planck Institute for the Physics of Complex Systems, 01187 Dresden, Germany}

\date{\today}

\begin{abstract}
We study rotating quasi-two-dimensional Bose-Einstein-condensates, in which atoms are dressed to a highly excited Rydberg state. This leads to weak effective interactions that induce a transition to a mesoscopic supersolid state. Considering slow rotation, we determine its superfluidity using Quantum Monte-Carlo simulations as well as mean field calculations. For rapid rotation, the latter reveal an interesting competition between the supersolid crystal structure and the rotation-induced vortex lattice that gives rise to new phases, including arrays of mesoscopic vortex crystals.
\end{abstract}


\maketitle
Superfluidity, i.e., the frictionless flow of a liquid, is one of the most spectacular manifestations of quantum mechanical behavior on a macroscopic scale. Generally, superfluidity can be characterized from the response of a many-body system to a slow, externally imposed rotation. A classical fluid enclosed in a rotating vessel will be dragged along and eventually rotate with it. A  superfluid, on the other hand, will, due to a lack of viscosity, remain stationary, and, for sufficiently fast rotation, form quantized  vortices that arrange on a regular lattice \cite{mcw00,arv01}. Superfluidity has long been speculated to occur even in solid states of matter, combining the seemingly antithetical qualities of crystalline order and non-dissipative flow \cite{al69,che70,leg70}. Recent experimental evidence for such a peculiar supersolid phase in $^4$He crystals \cite{kc04a,kc04b} has sparked an intense debate about its physical origin, which, however, remains controversial \cite{HeSuso}.

On the other hand, ultracold gases have emerged as a powerful laboratory to study a diverse range of many-body problems, including discrete supersolid states in optical lattices \cite{OL}. A promising route to the observation of \emph{continuous} supersolids has been recently laid out \cite{{hnp10,cjb10}}, based on off-resonant dressing of atomic Bose-Einstein condensates (BECs) to high-lying Rydberg states \cite{ssz00,pmb10,mhs11,mls09,wae11}. The effective atomic interactions resulting from such a Rydberg-dressing 
provide a clean realization of a simple model for supersolidity \cite{suso_model}. Given the range of available techniques for accurate probing of BECs \cite{bdz08} and recent advances in the manipulation of cold Rydberg atoms \cite{RydExp}, this approach holds promise for the observation of supersolidity under well-defined and highly controllable conditions.

\begin{figure}[t!]
\includegraphics[width=.95\columnwidth, angle=0]{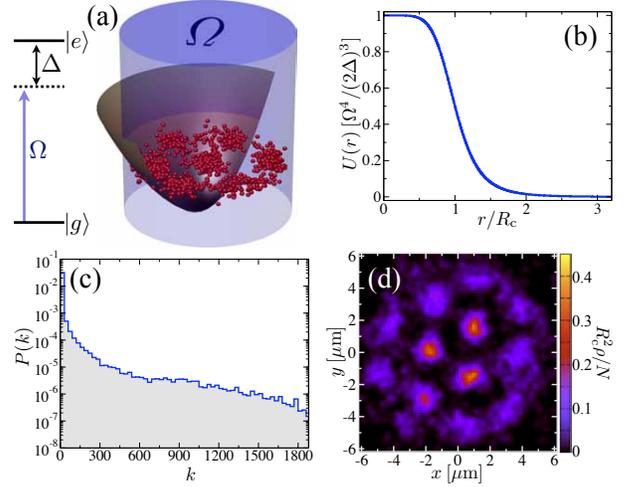}
\caption{\label{Fig1}(color online) (a) Schematics of the considered setup in which the groundstates ($|g\rangle$) of condensed atoms are off-resonantly coupled to high-lying Rydberg states ($|e\rangle$) with a large laser detuning $\Delta$ and a small Rabi frequency $\Omega\ll\Delta$. The resulting effective interaction, shown in (b), can lead to the formation of mesoscopic supersolids in a quasi two-dimensional condensate, exemplarily shown in panels (c) and (d). The extended plateau of the histogram $P(k)$ of $k$-particle permutations, shown in (c), indicates a large superfluid fraction, while the density shown in (d) demonstrates crystallization. The results correspond to $N=3000$ Rubidium atoms confined in a trap with $\omega_{\rm tr}/2\pi=125$Hz and dressed to Rb($50S_{1/2}$) Rydberg states with realistic laser parameters of $\Delta/2\pi=75$MHz and $\Omega/2\pi=2.2$MHz. The depicted QMC results yield a large superfluid fraction $f_{\rm s}=0.51(4)$ for an experimentally accessible temperature of $T=42$nK.}
\end{figure}

In this Letter, we explore the possibility for creating and detecting mesoscopic supersolids in confined, Rydberg-dressed BECs by probing their response to forced trap rotations. For infinitesimally small rotation frequencies, large-scale simulations  demonstrate that crystalline order and superfluidity (see Fig.\ref{Fig1}) persist at well-accessible temperatures, enabling supersolid creation in cold atom experiments. Combining the results from first-principle Quantum Monte Carlo simulations and mean field calculations, we reveal a universal behavior of the crystalline ordering and the superfluidity that enables simple estimates of the system behavior over a wide range of experimentally relevant conditions. At higher frequencies, the rotation induces drastic structural changes due to an interesting competition between the vortex lattice, controlled by short-range collisions, and the supersolid crystal arising from the finite-range Rydberg interactions.

A Rydberg-dressed BEC is a gas of Bose-condensed atoms whose ground state $|g\rangle$ is far off-resonantly coupled to a highly excited Rydberg state $|e\rangle$ with a Rabi frequency $\Omega\ll\Delta$ much less than the corresponding laser detuning $\Delta$ [see Fig.\ref{Fig1}(a)] \cite{ssz00,hnp10,cjb10,pmb10,mhs11}. Specifically, we consider two-photon coupling to $nS_{1/2}$ Rydberg states of alkalines which leads to an effective interaction \cite{hnp10}
\begin{equation} \label{pot}
 W(r) = \frac{\tilde C_6}{R_{\rm c}^6+r^6}
\end{equation}
between the dressed ground state atoms, where $\tilde C_6=\left(\tfrac{\Omega}{2\Delta}\right)^4C_6$, $R_{\rm c}=(C_6/2\hbar\Delta)^{1/6}$ and $C_6$ is the strength of the van-der-Waals interaction between Rydberg atoms \cite{sing05}. At large distances $r\gg R_{\rm c}$, the interaction resembles the Rydberg-Rydberg atom interaction $C_6/r^6$, suppressed by a factor $(\Omega/2\Delta)^4$ since only a small Rydberg state fraction $(\Omega/2\Delta)^2$ is admixed to the ground state. At small distances the dipole blockade \cite{blockade} prevents two-atom dressing, such that $W(r)$ saturates to a value of $\left(\tfrac{\Omega}{2\Delta}\right)^3\hbar\Omega$, independently of the addressed Rydberg state [see Fig 1(b)].

We consider quasi two-dimensional condensates \cite{phs00,gvl01}, that are strongly confined along the $z$-direction. Transversely, the atoms are weakly confined in a harmonic trap with a frequency $\omega_{\rm tr}$. Scaling lengths by the oscillator length $l=\sqrt{\tfrac{\hbar}{m\omega_ {\rm tr}}}$ ($m$ is the atomic mass) and frequencies by $\omega_ {\rm tr}$, we obtain the following  Hamiltonian in the rotating frame
\begin{eqnarray}\label{ham}
	\hat H &=& -\tfrac12\sum_i\tilde{\nabla}_i^2 + \tfrac12\sum_i \left( \tilde{r}_i^2  - \Omega\tilde{L}_{z,i} \right) \nonumber\\
	 &&+ \sum_{i<j}\frac{\alpha}{r_c^6+|\tilde{{\bf r}}_{i}-\tilde{{\bf r}}_{j}|^6}+ \gamma\delta(\tilde{{\bf r}}_{i}-\tilde{{\bf r}}_{j})  ,
\end{eqnarray}
where $\gamma = \sqrt{8\pi}a_{\rm s}/l_z$ is the effective strength of the contact interaction due to s-wave collisions with scattering length $a_{\rm s}$, $\alpha=m\tilde{C}_6/(\hbar^2 l^4)$, $r_c=R_c/l$, $\Omega$ is the rotation frequency in units of the trapping frequency, $\tilde{{\bf r}}_{i}={\bf r}_i/l$ is the scaled position of the $i$th atom and $\tilde{L}_{z,i}$ denotes the corresponding angular momentum operator.

Focussing on the effects of the Rydberg interactions we first assume $\gamma=0$ and investigate the stationary ($\Omega=0$) equilibrium thermodynamics of $\hat{H}$ by Quantum Monte Carlo (QMC) simulations \cite{cep95} based on the continuous-space Worm algorithm \cite{bps06}. Fig.\ref{Fig1}(d) shows the computed density profile for a large number ($N=3000$) of  Rubidium atoms at a temperature $T=42$nK coupled to Rb($50S_{1/2}$) Rydberg states with $\Omega/2\pi=2.2$MHz and $\Delta/2\pi=75$MHz. The formation of a self-assembled droplet crystal is evident. Note that the experimentally relevant parameters correspond to very weak interactions $k_{\rm B}^{-1}W(r)<k_{\rm B}^{-1}\hbar\Omega^4/(2\Delta)^3\sim10^{-1}{\rm nK}\ll T$. In this weak coupling limit, crystallization, nevertheless, occurs due to the presence of many atoms in each blockade sphere, forming a solid of liquid droplets with collectively enhanced interactions. On the other hand, the average number of Rydberg excitations in each droplet is only $\sim0.04$, which diminishes Rydberg state decay \cite{hnp10,pmb10} and renders many-body effects, that tend to counteract the binary interaction (\ref{pot}) \cite{hwp10}, unimportant.

In addition, the QMC simulation yields a large superfluid fraction of $f_{\rm s}\approx0.5$, showing that the observed mesoscopic crystal is indeed a supersolid. $f_{\rm s}$ corresponds to the fraction of atoms that decouples from a vanishingly small trap rotation, such that $f_s=1-I_{qm}/I_c$, where $I_c$ and $I_{qm}$ are the classical and the actual (i.e. quantum mechanical) moments of inertia, that can be computed within QMC \cite{skc89}. The finite value of $f_s$ is also indicated in Fig.\ref{Fig1}(c) by the extended plateau of the probability $P(k)$ that $k$ particles exchange within the system. $P(k)$ is finite for all $k \leq N$ in a superfluid, but drops to zero in a classical crystal \cite{krauth2006}.

The large superfluid fraction arises from the extended inner plateau of the interaction potential (\ref{pot}) which can accommodate a large number of atoms that maintain a sizable superfluid fraction, even in the defect-free droplet crystals discussed in this work. As a result, the superfluidity does not critically depend on the size and specific structure of the mesoscopic crystals. This behavior is in contrast to mesoscopic assemblies with pure power-law potentials \cite{powerlaw} where the superfluidity depends strongly on the cluster geometry and vanishes in the bulk-limit of infinite defect-free crystals \cite{defects}. 

\begin{figure}[t!]
\includegraphics[width=.99\columnwidth, angle=0]{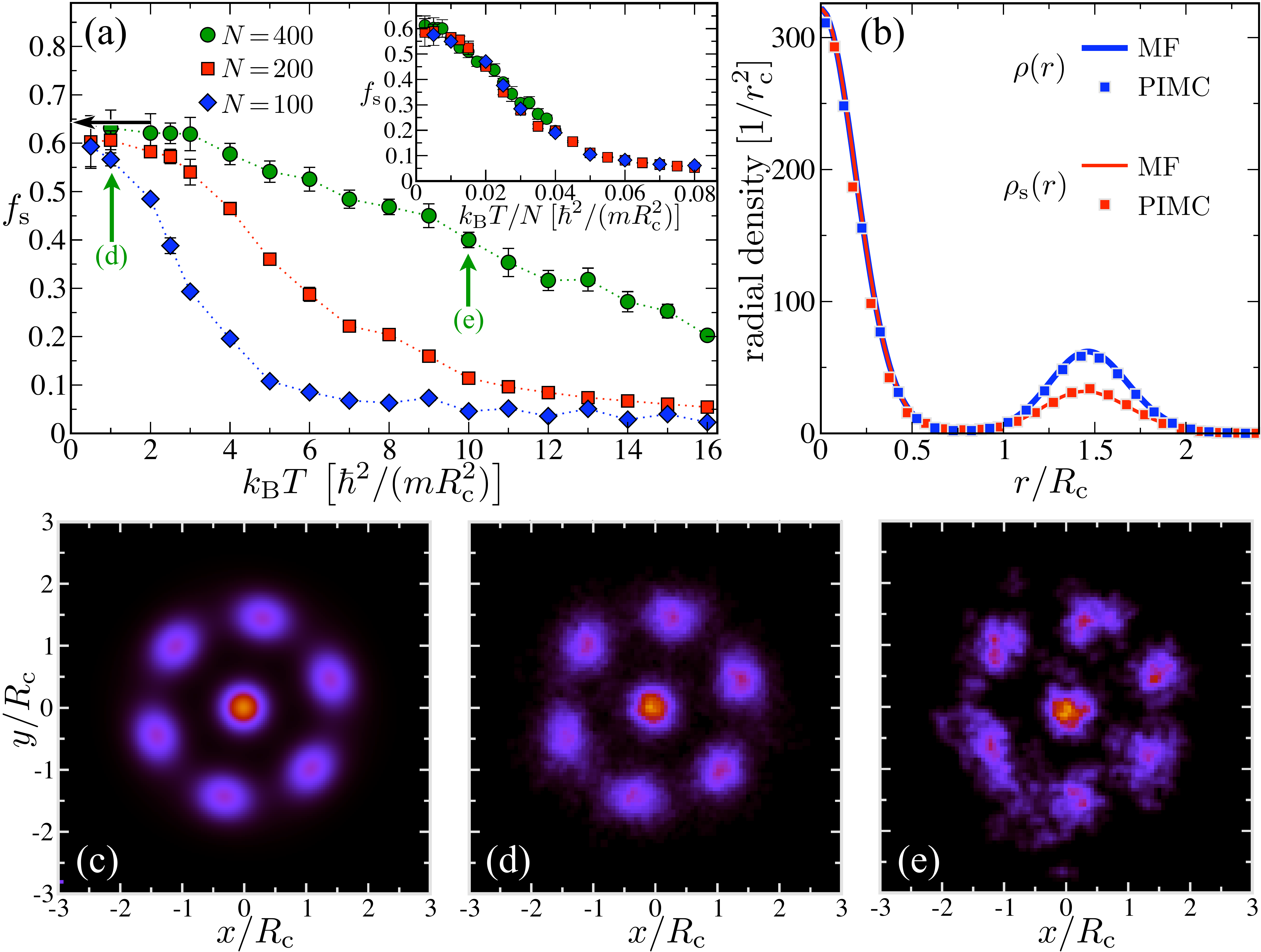}
\caption{\label{Fig2} (color online) (a) Temperature dependence of the superfluid fraction for different particle numbers and $\alpha N=24500$, $\gamma=0$ and $r_c=2.65$. The horizontal arrow indicates the $T=0$ MF prediction [eq.(\ref{sf_MF})]. As shown in the inset, all data points collapse on a single curve upon scaling $T$ by $N$. The vertical arrows indicate the temperatures used in (d) and (e). (b) Comparison of the radial density $\rho$ and superfluid density $\rho_{\rm s}$ for $N=400$. Panels (c)-(e) show the corresponding ground state densities obtained from (a) the MF approximation eqs.(\ref{GPE}) and (\ref{sf_MF}) and (b,c) the QMC simulations.}
\end{figure}

The temperature dependence of $f_{\rm s}$ is shown in Fig.\ref{Fig2}(a) for different atom numbers $N$ and interaction strengths $\alpha$, but with the product $\alpha N={\rm const.}$ held constant. Apparently, superfluidity extends to higher temperatures for larger values of $N$, which is readily understood from simple arguments. As will be explained below, the crystal structure remains unchanged for $\alpha N={\rm const.}$, such that increasing $N$ merely increases the number of atoms in each droplet. As a result, the system becomes less susceptible to fluctuations, which facilitates exchange between the mesoscopic crystal sites and, hence, leads to larger $f_{\rm s}$. 
Upon rescaling the temperature with $N$, all data collapse on a single curve [inset of Fig.\ref{Fig2}(a)], which is consistent with the linear density scaling of the critical temperature for the Berezinskii-Kosterlitz-Thouless transition \cite{bkt} in the weak-coupling limit \cite{fh88}. This universal behavior of the crystalline order and superfluidity with respect to $\alpha N$ and $T/N$ can be used to predict the system behavior for a wider range of parameters. For instance, at a typical interaction range $R_{\rm c}=2\mu$m and a large number of $N=10^4$ Rb atoms, the calculations of Fig.\ref{Fig2} imply a crystal structure as in Fig.\ref{Fig2}(e) with a superfluid fraction of $f_{\rm s}\approx0.4$, both persisting at a large temperature of $T\approx350$nK.

The local superfluid density $\rho_{\rm s}$ is also closely linked to slow trap rotations, and can be obtained within QMC from a local area-estimator \cite{kpw06} of the non-classical rotational inertia. In his seminal paper on supersolidity in Helium, Leggett \cite{leg70} proposed a far simpler approach to estimate the superfluid density by considering the response of a liquid in a slowly rotating ring-shaped container. Within a mean field (MF) approximation, simple phase matching arguments yield the radial superfluid density \cite{leg70}
\begin{equation} \label{sf_MF}
\rho_{\rm s}(r)= \frac{2\pi}{\int {\rm d}\varphi\rho({\bf r})^{-1}}\;,
\end{equation}
solely in terms of the single-atom density $\rho({\bf r})$, integrated over the polar angle $\varphi$. This expression provides an upper bound of the exact superfluid density $\rho_{\rm s}(r)$ and offers an intuitive picture for the superfluid  properties of the found droplet crystals, suggesting that the maintenance of a superfluid flow is foremost hampered by regions of low density \cite{leg70,sas76,gbl11} . With increasing interactions the density modulations become more pronounced such that $\rho_{\rm s}(r)$ eventually vanishes, since the atoms can not pass through extended regions where $\rho(r,\varphi)\rightarrow0$. In this strong-coupling limit, that goes beyond the MF approximation, QMC simulations show that the system enters a globally insulating phase \cite{cjb10}. Likewise, eq.(\ref{sf_MF}) gives $\rho_{\rm s}(r)=\rho(r)$ in the non-modulated phase, i.e. a perfect superfluid.

In order to test this intuitive picture we also performed MF calculations, based on the corresponding nonlocal Gross-Pitaevskii equation (GPE)
\begin{eqnarray}\label{GPE}
i\partial_t\tilde{\psi}(\tilde{{\bf r}})&=&\left[-\frac{\nabla^2}{2} + \frac{\tilde{r}^2}{2} - \Omega\tilde{L}_z\right.\\
&&+\left.\gamma N |\tilde{\psi}(\tilde{{\bf r}})|^2+ \alpha N\int {\rm d}\tilde{{\bf r}}^{\prime}\frac{|\tilde{\psi}(\tilde{{\bf r}}-\tilde{{\bf r}}^{\prime})|^2}{r_c^6+\tilde{r}^{\prime6}}\right]\tilde{\psi}(\tilde{{\bf r}}) \nonumber
\end{eqnarray}
where $\tilde{\psi}(\tilde{{\bf r}})$ denotes the normalized ($\int|\tilde{\psi}|^2{\rm d}\tilde{{\bf r}}=1$) condensate wavefunction and yields the single atom density $\rho({\bf r})=N|\tilde{\psi}(\tilde{{\bf r}})|^2$. Eq.(\ref{GPE}) predicts that within MF the crystal structure depends only on the product $\alpha N$, as found above from the QMC simulations.
At very low temperatures, the MF calculations yield virtually the same crystal structure as the QMC results [cf. Fig.{\ref{Fig2}(c) and (d)]. In this zero-temperature limit the superfluidity approaches a constant value independent of the atom number $N$ [see Fig.\ref{Fig2}(a)], with $\alpha N$ held constant. This behavior can also be understood from eqs.(\ref{sf_MF}) and (\ref{GPE}), which depend solely on the product $\alpha N$. Indeed, the radial atomic density and superfluid densities obtained from eqs.(\ref{sf_MF}) and (\ref{GPE}) closely reproduce the QMC results, demonstrating that eq.(\ref{sf_MF}) not only provides an upper bound \cite{leg70,gbl11} of the superfluidity, but yields an accurate description of the structural and superfluid properties for the present realization of a supersolid.

\begin{figure*}
\begin{center}
\includegraphics[width=.99\textwidth]{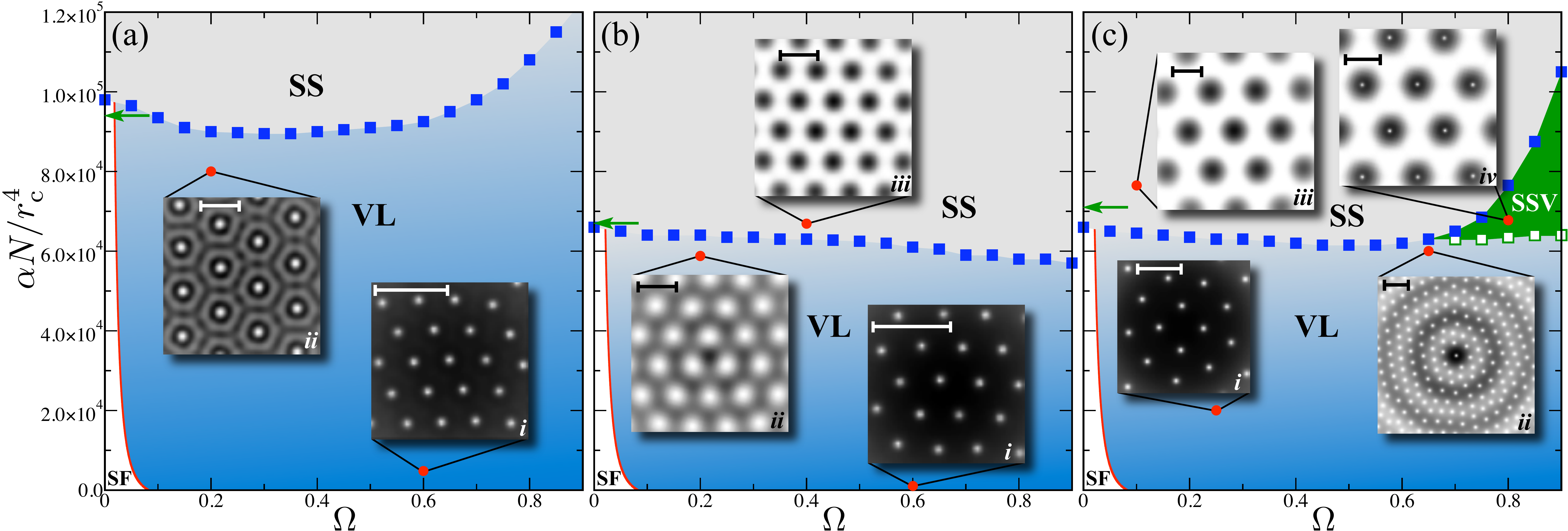}
\caption{\label{Fig3}(color online) ``Phase diagrams`` for $\gamma N=10^4$ and (a) $r_c=1$, (b) $r_c=3$, (c) $r_c=5$. The insets display ground state density profiles at the indicated  parameters, showing only the inner trap region for better visibility, where dark shading indicates high density and vice versa. The dimension of each inset is indicated by the vertical bar corresponding to a length of $5l$. The horizontal arrows at $\Omega=0$ show the supersolid transition obtained from LDA estimates of the roton instability as described in the text.}
\end{center}
\end{figure*}

Having established the predictive power of the MF approximation for $T\rightarrow0$, we can apply eq.(\ref{GPE}) to rotating assemblies.
Previous MF studies of rotating dipolar condensates \cite{crs05} have revealed a range of different vortex lattices as a function of the dipolar interaction strength.
In the present case, the ground state depends on four parameters $r_c$, $\Omega$, $\gamma N$ and $\alpha N$, which makes a complete parameter scan prohibitively demanding. Being primarily interested in the effects of the Rydberg interactions and focussing on the large-$N$ regime, we set $\gamma N=10^4$ \cite{NaBEC}. In order to scan the remaining parameter space, we use the following procedure: For a given set of $\Omega$ and $r_c$ we first calculate the ground state for $\alpha=0$ via complex time evolution, which agrees with the results \cite{tku02} for contact interactions. Subsequently, we slowly increase $\alpha N$ and follow the complex time evolution \cite{hnp10,mhs11,tku02} of the BEC. This yields the ``phase diagrams`` depicted in Fig.\ref{Fig3}.

In the region of small $\alpha N$ and $\Omega$ (SF in Fig.\ref{Fig3}) we find a simple superfluid, with $\rho_{\rm s}=\rho$ and an unstructured density profile.  Upon increasing the rotation frequency above a critical value, vortices start to form and eventually arrange on a regular lattice, whose spacing decreases for increasing rotation frequency and interaction strengths (VL in Fig.\ref{Fig3}). For small values of $\alpha N$ the nonlocal character of the corresponding interaction in eq.(\ref{GPE}) is of minor importance, such that the total interaction can be described via an effective contact term $\gamma^{\prime}=\gamma+2\pi^2\alpha/(3^{3/2}r_{\rm c}^4)$.
In this regime, we indeed find ordinary vortex lattices determined by $\gamma^{\prime}N$ [see Figs.\ref{Fig3}(a){\it i}, (b){\it i} and (c){\it i}]. With increasing $\alpha N$ the finite-range  of the dressing-induced interaction becomes important and changes the symmetry of the vortex lattice. For small $r_{\rm c}$ [Fig.\ref{Fig3}(a)] the system is already close to the bulk limit, where one finds a transition to a honeycomb density patterns. For large $r_{\rm c}$ finite size effects become dominant and lead to the formation of concentric rings superimposed on the underlying vortex lattice [Fig.\ref{Fig3}(c){\it ii}].

The finite system size also affects the crossover to supersolid states \cite{contrast}. In the bulk limit, the critical interaction strength $\alpha_{\rm 2D}N$ can be linked to a roton instability of the unmodulated groundstate \cite{hnp10}, and, therefore, decreases with increasing rotation frequency (at larger $\Omega$ it increases again due to the decreasing density caused by the centrifugal potential). For small $r_{\rm c}$ [Fig.\ref{Fig3}(a)], the critical value at $\Omega=0$ can, thus, be estimated from a local density approximation (LDA) of the roton instability, which gives $r_c^{-4}\alpha N|\tilde{\psi}({\bf 0})|^2=31.9+6.1\gamma$ and for the parameters of Fig.\ref{Fig3}(a), $r_c^{-4}\alpha N=9.3\times10^4$. For larger $r_{\rm c}$ the supersolid transition is preceded by the aforementioned ring formation. Assuming that the transition is initiated by a one-dimensional roton instability of the innermost ring one obtains $r_c^{-4}\alpha_{\rm 1D} N=6.7\times10^4$ and $r_c^{-4}\alpha_{\rm 1D} N=7.2\times10^4$ for the parameters of Fig.\ref{Fig3}(b) and (c), respectively, in good agreement with the numerics.

\begin{figure}[b]
\begin{center}
\includegraphics[width=.99\columnwidth]{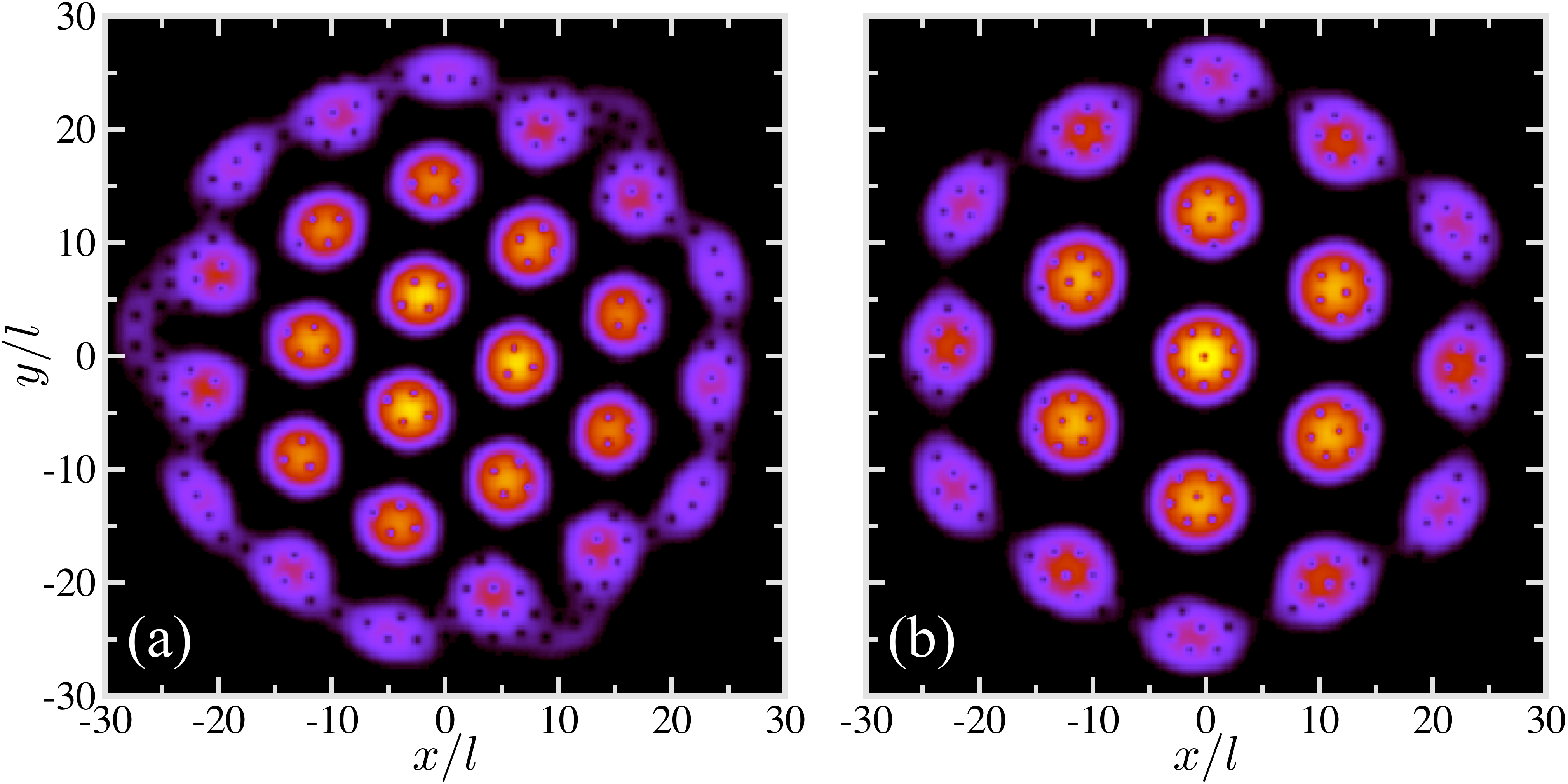}
\caption{\label{Fig4} Density for $\Omega=0.8$ and (a) $r_{\rm c}=7$, $\alpha N/r_c^4=7.5\times10^4$ and (b) $r_c=9$, $\alpha N/r_c^4=9\times10^4$.  }
\end{center}
\end{figure}

The supersolid phase (SS in Fig.\ref{Fig3}) retains the vortices. They are, however, pushed into the low-density regions and form a honeycomb vortex lattice that does not affect the structure of the supersolid droplet crystal (see Figs.\ref{Fig3}(b){\it iii} and (c){\it iii}). At larger $r_{\rm c}$, on the other hand, there is an interesting competition of length scales between the supersolid crystal and the vortex lattice. For sufficiently high rotation frequencies the vortex density, and, hence, the vortex-vortex interaction, exceeds a critical value at which it becomes energetically favorable to form vortices inside the superfluid droplets [SSV in Fig.\ref{Fig3}(c)]. As shown in Fig.\ref{Fig3}(c){\it iv}, the crystal structure of this additional vortex lattice is imposed by the triangular supersolid. The minimum rotation frequency, to support these states increases as $r_{\rm c}$ is reduced. Since the confinement ceases for $\Omega\ge1$, this implies a minimum interaction range $r_{\rm c}^{\rm min}(\gamma N)$ in order to observe the SSV states for a given contact interaction $\gamma N$. For larger $r_{\rm c}$ the on-site number of vortices successively increases, which eventually form a mesoscopic  crystal of small vortex crystals as shown in Fig.\ref{Fig4}.

In this work, we demonstrated novel vortex transitions in a realistic model of supersolids in the regime of weak interactions and large atom numbers per blockade sphere, which is most directly accessible to experiments. Extending the present study to the regime of strong coupling, future work may address how strongly correlated supersolids can be stabilized in finite-temperature cold atom experiments, which will more closely resemble the physical mechanism thought to underlie supersolidity in He. Along these lines, the realization of the found vortex crystals in the regime of strong interactions and small particle numbers will open the way towards studying quantum Hall-like phenomena in rapidly rotating self-assembled structures. This will require a beyond-MF description of the correlated many-body dynamics induced by rapid rotation.

We thank P. Zoller and M. Boninsegni for valuable discussions. This work was supported by MURI, AFOSR, EOARD, IQOQI, the EU through NAME-QUAM, COHERENCE and AQUTE.


\begin{thebibliography}{99}
\bibitem{contrast} The transition to a mesoscopic supersolid state is signaled by an abrupt increase of the density contrast $c=1-\rho_{\rm max}/\bar{\rho}$, which we determine from the maximum particle density $\rho_{\rm max}$ and the average density $\bar{\rho}$ of the atomic cloud.
\end{thebibliography}
\end{document}